\documentclass[twocolumn,preprintnumbers,amsmath,amssymb,superscriptaddress]{revtex4-1}

\usepackage{graphicx}
\usepackage{float}
\usepackage{bm}
\usepackage{color}
\usepackage{ulem}

\begin{document}

\title{Single shot measurement of the photonic band structure in a Floquet-Bloch lattice realised with coupled fiber rings}

\author{Corentin Lechevalier}
\affiliation{Univ. Lille, CNRS, UMR 8523 - PhLAM -
  Physique des Lasers Atomes et Mol\'ecules, F-59 000 Lille, France}
\author{Cl\'ement Evain}
\affiliation{Univ. Lille, CNRS, UMR 8523 - PhLAM -
  Physique des Lasers Atomes et Mol\'ecules, F-59 000 Lille, France}
\author{Pierre Suret}
\affiliation{Univ. Lille, CNRS, UMR 8523 - PhLAM -
  Physique des Lasers Atomes et Mol\'ecules, F-59 000 Lille, France}
\author{Fran\c{c}ois Copie}
\affiliation{Univ. Lille, CNRS, UMR 8523 - PhLAM -
  Physique des Lasers Atomes et Mol\'ecules, F-59 000 Lille, France}
\author{Alberto Amo}
\affiliation{Univ. Lille, CNRS, UMR 8523 - PhLAM -
  Physique des Lasers Atomes et Mol\'ecules, F-59 000 Lille, France}
\author{St\'ephane Randoux}
\email{stephane.randoux@univ-lille.fr}
\affiliation{Univ. Lille, CNRS, UMR 8523 - PhLAM -
  Physique des Lasers Atomes et Mol\'ecules, F-59 000 Lille, France}

\begin{abstract}
  Floquet-Bloch lattices are systems in which wave packets are subjet to periodic modulations both in time and space, showing rich dynamics. While this type of lattices is difficult to implement in solid-state physics, optical systems have provided excellent platforms to probe their physics: among other effects, they have revealed genuine phenomena such as the anomalous Floquet topological insulator and the funnelling of light into localised interface modes. Despite the crucial importance of the band dispersion in the photon dynamics and the topological properties of the lattice, the direct experimental measurement of the Floquet-Bloch bands has remained elusive. Here we report the direct measurement of the Floquet-Bloch bands of a photonic lattice with a single shot method. We use a system of two coupled fibre rings that implements a time-multiplexed Floquet-Bloch lattice. By Fourier transforming the impulse response of the lattice we obtain the band structure together with an accurate characterization of the lattice eigenmodes, i. e. the amplitudes and the phases of the Floquet-Bloch eigenvectors over the entire Brillouin zone. Our results open promising perspectives for the observation of topological effects in the linear and nonlinear regime in Floquet systems.
\end{abstract}

\maketitle

\section{Introduction}\label{Sec1}

Optical waveguide arrays represent a class of periodic structures that has been the focus of intense research in the two last decades \cite{Garanovich:12,Christodoulides2003}. Such systems enabled direct observation with light of many fundamental lattice phenomena such as Bloch oscillations \cite{Morandotti:99,Pertsch:99,Wimmer:15}, Anderson localization \cite{Pertsch:04,Lahini:08,Vatnik:17}, discrete solitons \cite{Eisenberg:98,Mandelik:04}, and many others. Coupled waveguides have also been used to implement Floquet lattices in which photon wavepackets are subject to a periodic time modulation~\cite{Rechtsman2013b, Maczewsky2017, Mukherjee2017a, Bellec2017}. However, material constraints have limited the number of Floquet periods accessible in this configuration. Recently, discrete mesh lattices in time-multiplexed loop arrangements have been used to study elaborate Floquet-Bloch Hamiltonians with access to hundreds of both lattice sites and modulation periods \cite{Regensburger:11}. So far, time-multiplexed fiber loop schemes have demonstrated a high degree of flexibility \cite{Chalabi:19,Chalabi:20} and they have been employed to investigate a number of effects such as parity-time symmetry \cite{Miri:12,Regensburger:12} or topological control of light propagation \cite{Chen:18,Weidemann:20} to cite a few. Note that photonics now offers a number of other possibilities to design and engineer the so-called synthetic mesh lattices, see ref. \cite{Yuan:18} for a recent review.

Time-multiplexed photonic mesh lattices are often implemented by connecting two appropriately designed fiber loops with a directional $50/50$ fiber coupler \cite{Wimmer:17,Wen:20,Muniz:19}. In these systems the phenomenon of discrete diffraction created at the fiber coupler results in the spreading of light wavepackets across the whole array as Floquet–Bloch waves \cite{Bisianov:19,Gomez:13}. The linear propagation of wavepackets in the Floquet-Bloch lattices is determined by their photonic band structure. Despite the importance of this band structure, its experimental determination is not an obvious task. In previous experimental works, the dispersive properties of the lattices have been deduced from the measurement in space and time of the group velocity and of the broadening experienced by many individual light wavepackets \cite{Miri:12,Wimmer:15,Wimmer:17}. Recently, the measurement of the band structure in a different configuration based on the longitudinal optical modes of a single fibre ring has enabled the observation of synthetic spin-orbit coupling and quantum Hall ladders~\cite{Dutt2019, Dutt2020}.

In this paper, we report the experimental implementation of a method that allows the single-shot recording of the photonic band structure characterizing a Floquet-Bloch lattice realised with two coupled fiber loops. The experiment relies on the idea that the dispersive band structure of the lattice can be determined from the Fourier transform of its impulse response. The simultaneous measurement of the phase and of the amplitude of the impulse response of the lattice is achieved using a heterodyne technique. In addition to providing the dispersive band structure of the lattice (i.e. the dispersion relation connecting the quasi-energy and the Bloch momentum), our method provides the full and accurate characterization of the lattice eigenmode structure, i. e. the amplitudes and the phases of the Floquet-Bloch eigenvectors over the entire Brillouin zone. Let us notice that the idea that the eigenmode structure of a photonic mesh lattice can be determined using optical heterodyne technique has been proposed in Ref. \cite{Tikan:17}. To the best of our knowledge, this idea has however never been implemented in practice and the fact that the full eigenmode structure of the photonic mesh lattice can be determined from the measurement of its impulse response has not been considered before our work.

\section{The lattice model and the associated photonic band structure}\label{Sec:Model}

As shown in Fig. 1(a), we consider a system of two coupled fiber loops which is conceptually identical to those considered in previous experimental works~\cite{Regensburger:11, Wimmer:17}. The two fiber loops have an imbalanced path length $\Delta L=L_2-L_1$ which is chosen to be much shorter than the length $L_1$ (resp. $L_2$) of the short (resp. long) ring. When an optical pulse is injected in one of the loops, it is divided into two pulses after the coupler. These two pulses propagate along the short and long fiber rings before being split again at the fiber coupler. As discussed in details in ref. \cite{Wimmer:17,Wen:20} the dynamical evolution of the light pulses in this optical fiber system can be mapped onto the lattice shown in Fig. 1(b). Each round trip of the pulses in the rings represents a time step, labelled by the integer $m$, while the separation of pulses within a time step can be mapped into the pseudo-real space position of the lattice, labelled $n$. The time scale associated to the real space position is given by $\Delta L$, while the time step is determined by $(L_{1}+L_{2})/2$. The large difference between these two time scales permits a clear observation of the evolution of pulses at each time step at the ouput of the setup. 

The space-time evolution of the complex amplitude of light pulses in the lattice shown in Fig. 1(b) is commonly described using a simple set of two coupled algebraic equations~\cite{Wimmer:15,Wimmer:17}:
\begin{align}
u_n^{m+1}=\frac{1}{\sqrt{2}} (u_{n+1}^m +i \, v_{n+1}^m) \, e^{i \Phi(m)} \,  \label{U}, \\ 
v_n^{m+1}=\frac{1}{\sqrt{2}} (v_{n-1}^m +i \, u_{n-1}^m) \,.  \label{V} 
\end{align}

\noindent $u_n^{m}=u(n,m)$ (resp. $v_n^{m}=v(n,m)$) represents the complex amplitude of the pulses in the short (resp. long) loop at the $n$th position in the pulse train and at the $m$th round trip (time step) in the fiber loop system\cite{Wimmer:15,Wimmer:17}. $\Phi(m)$ is an extra phase gained by the pulses in the shorter (U) ring thanks to the addition of a phase modulator (PM in Fig. 1(a)). Its value changes sign at each time step ($\Phi(m)=(-1)^{m+1} \, \phi$), as depicted in Fig. 1(b). 

Following Ref. \cite{Wimmer:15}, the Floquet-Bloch mode eigenstates $(U,V)^T$ of the lattice are obtained by decomposing $u_n^{m}$ and $v_n^{m}$ on a discrete basis of Fourier modes
\begin{align}\label{eigenmodes}
    \begin{bmatrix} 
           u_{n}^{m} \\
           v_{n}^{m} \\
    \end{bmatrix}
  = \begin{bmatrix}  
           U \\
           V \\
  \end{bmatrix}
  e^{i \frac{Q n}{2}}  e^{i \frac{\theta m}{2} }.
\end{align}
This means that in reciprocal (Fourier) space, the variables $n$ and $m$ are conjugated with the ``Bloch momentum'' $Q$ and the ``quasi-energy'' $\theta$, respectively (see e. g. ref. \cite{Wen:20,Wimmer:17}).

Substituting Eq. (\ref{eigenmodes}) into Eqs (\ref{U}), (\ref{V}), it can be easily shown that the dispersion relation of the system presents two bands that are periodic both along the quasi-energy and momentum dimensions and that are given by \cite{Wimmer:17}:
\begin{align}
\label{eq:disp}
\cos\theta = \frac{1}{2} (\cos Q - \cos\phi).
\end{align}
As shown e.g. in ref. \cite{Wimmer:17}, one of the great advantages of the photonic lattice described in Fig. 1  is that the band structure can be easily modified by varying the value of $\phi$.

The two bands are represented in Fig.~1(c) as a function of the parameter $\phi$. For $\phi=0$ (equivalently, $\phi= 2\pi$, blue lines in Fig.~2(c)), the lattice model has two bands that are gapped at the center of the Brillouin zone and touch at the edges due to the periodicity in quasi-energy. For other values of $\phi$ the bands can be fully gaped (red line) or touch in the center of the Brillouin zone (green line at $\phi=\pi$). Considering a light wavepacket having a well-defined mean Bloch momentum $Q_0$ together with a narrow momentum spread $\Delta Q$, the group velocity of this wavepacket in the $(n, m)$ (space-time) representation space is determined by the local slope of the excited band while the local curvature of the band determines the dispersive broadening of the wavepacket in space and time \cite{Wen:20,Wimmer:17}.

\begin{figure}[t!]
  \includegraphics[width=\linewidth]{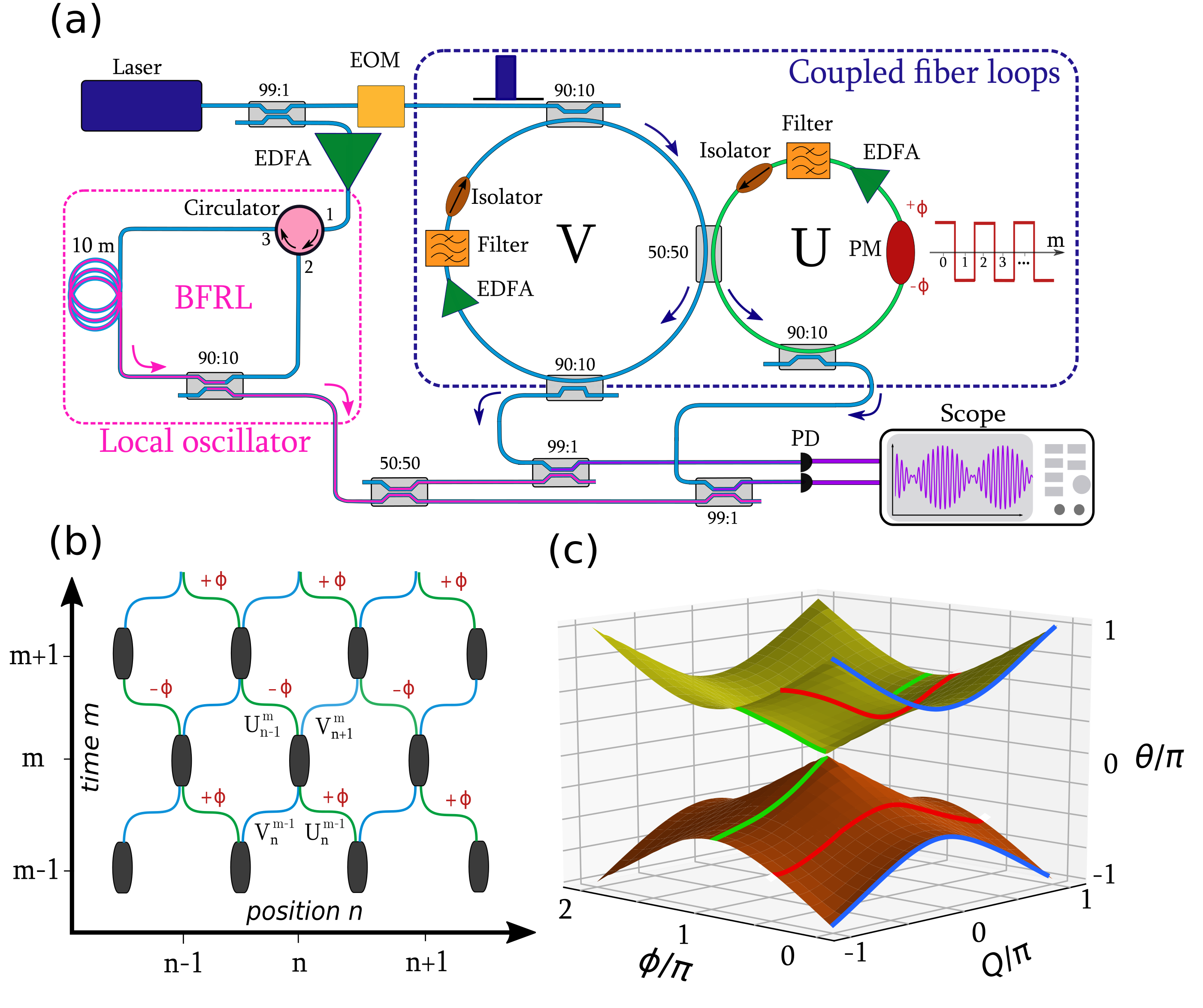}
  \caption{(a) Experimental setup showing the system of two coupled loops made with polarization maintaining fibers (PMFs) having lengths $L_1=30.27$ m and $L_2=30.72$ m. The local oscillator used for the heterodyne measurement is a Brillouin fiber ring laser (BFRL) that is frequency detuned by $\sim 10.8$ GHz from the frequency of the signal circulating inside the loops. (b) Schematic represention of the Floquet-Bloch lattice on which the evolution of the light pulses circulating inside the loops can be mapped. (c) Photonic band structure (Bloch momentum $Q$ versus quasi-energy $\theta$) for different values of the phase modulation $\phi$ within the U loop. 
  }
\end{figure}

\begin{figure*}[!ht]
  \centering
  \includegraphics[width=1.\linewidth]{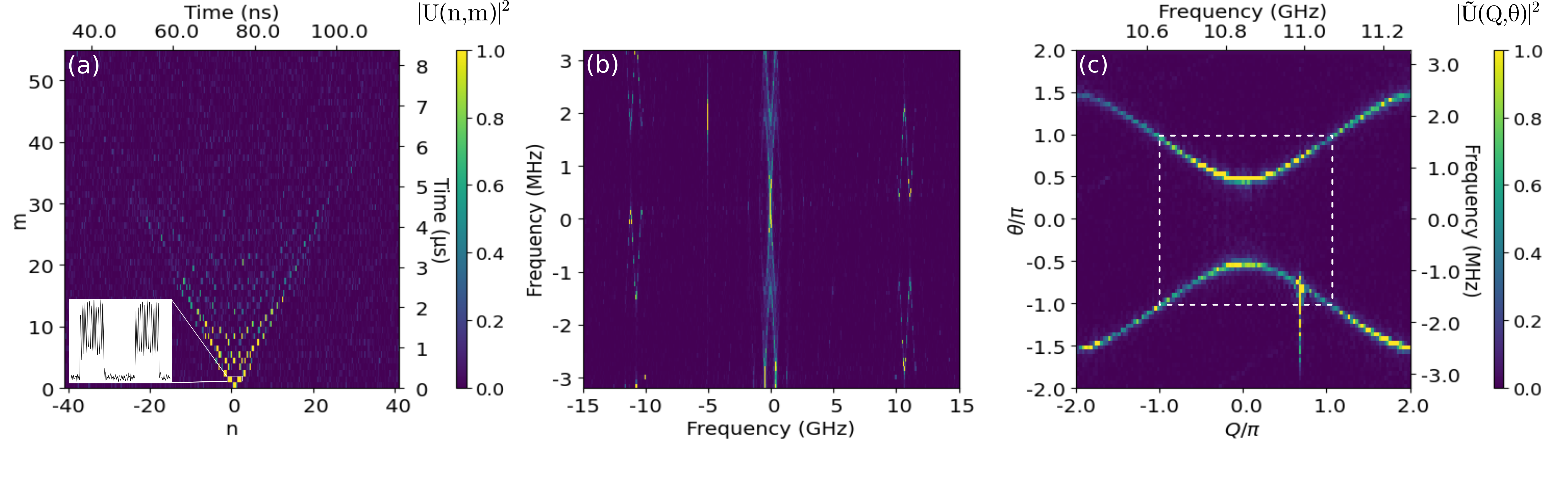}
  \caption{Experimental results showing (a) the space-time evolution measured in the U loop when a short pulse is injected in the V loop. The impulse response of the lattice shown in (a) is beated against a reference laser field. This gives square pulses being modulated at $\sim 10.8$ GHz, as shown by the inset in (a). The fringe pattern expanding in space ($n$) and time ($m$) is numerically Fourier transformed in the two dimensions. This gives the 2D Fourier power spectrum plotted in (b) with one central component and two weak side bands at $\sim \pm 10.8$ GHz. The 2D Fourier spectrum plotted in (c) represents a zoomed view of the spectrum shown in (b) between $10.45$ GHz and $11.33$ GHz. In terms of normalized units ( ``Bloch momentum'' $Q$ and ``quasi-energy'' $\theta$), the photonic band structure of the lattice is measured for $Q \in [-2 \pi, 2 \pi]$, $\theta \in [-2 \pi, 2 \pi]$. The first Brillouin zone is depicted by the square region plotted in (c) with white dashed lines.  }
\end{figure*}

\section{Measurement of the photonic band structure}\label{Sec2}

Our experimental determination of the band structure characterizing a Floquet-Bloch lattice is based on the single-shot simultaneous measurement of the phase and amplitude of the impulse response of the lattice. This is achieved using a heterodyne measurement where the wavefield at the output of the double loop system is beated against a reference field which is detuned from the frequency of the wavefield by $\sim 10.8$ GHz.

As shown in Fig. 1(a), a single-frequency laser at $1550$ nm is splitted into two arms by using a $99/1$ fiber coupler. The high intensity arm is directed towards the double ring system. A short square pulse with a duration of $1$ ns and a peak power of $\sim 6$ mW is produced using an electro-optic modulator (EOM) before being injected into the two coupled fiber loops. Each fiber loop incorporates a narrow-bandwidth optical filter, an optical isolator and an Erbium-doped fiber amplifier (EDFA) to partially compensate all round-trip losses. The length $L_1$ of the shorter loop is $30.27$ m and the difference in loop lengths is $\Delta L=0.45$ m. With these values the average round trip time is $\bar{T}=(2L_1+\Delta L)/(2 v) \simeq 152.5$ ns and the time difference between the two loops is $\Delta T=\Delta L/v \simeq 2.26$ ns, $v$ being the velocity of light in the fiber at $1550$ nm. The choice of these values for the ring lengths is one of the key features of our experiment: the fiber ring lengths are more than one order of magnitude smaller than in other similar experimental setups \cite{Wen:20,Wimmer:17} to keep the whole physical distance covered by pulses circulating inside the loops smaller than the coherence length of the local oscillator delivering the reference field.

The local oscillator that delivers the reference field used in the hererodyne measurement is a Brillouin fiber ring laser (BFRL, pink dotted rectangle in Fig.~1(a)). A small part of the laser field extracted before the EOM is amplified at the Watt level using an EDFA and it is used as a pump field for the BFRL. The BFRL delivers a Stokes field with is frequency-downshifted with respect to the pump frequency by $\sim 10.8$ GHz, the frequency of acoustic waves propagating inside the optical fiber \cite{Smith:91}. It is well known that BFRLs deliver a Stokes radiation having a linewidth much narrower that the one of their pump laser \cite{Debut:00,Geng:06}. This feature, recently exploited for the improved operation of atomic clocks \cite{Loh:20}, is used here to achieve a coherent beating between the narrow-bandwidth Stokes field and the pulses that propagates over the kilometric range associated with the tens of round trips made within the two coupled fiber loops.

Note that the fiber setup schematically shown in Fig. 1(a) is fully made with polarization maintaining fibers (PMFs) and with PMF components. The laser field has a linear polarization state with a power extinction ratio better than $1:100$ both in the the BFRL and in the double loop system. The fact that the light polarization state does not fluctuate in the fiber system has the advantage to maximize the contrast of the beating signal between the local oscillator and the pulses at the output of the loop system. Moreover the experimental results can be compared adequately with theoretical results provided by scalar models where vectorial (polarization) effects are ignored.

Figure 2(a) shows the spatio-temporal evolution measured at the output of the U ring after injection of a single initial pulse in the $V$ ring and in the absence of phase modulation ($\Phi(m)=0$). The sequence of measured pulses shows trains of pulses separated by $\bar{T}$. We use this time to order the pulses as a function of time step $m$ (round trip number) in Fig.~2(a), see Supplementary material for details about the reconstruction of the space-time diagram from the recorded time signal. The output of the U ring is combined in a $99:1$ beamsplitter with the BFRL local oscillator. A fast photodiode connected to a fast oscilloscope is used to record the beat signal between the output of the coupled loop system and this oscillator, as shown in Fig.~1(a). In this way, each individual pulse at the output of the loops has its amplitude that is fastly modulated in time at the beat frequency of $\sim 10.8$ GHz, see inset in Fig. 2(a) that shows the measured amplitude of the train of pulses at time step $m=2$. The evolution of the relative phase between the light pulses within each round trip and between different round trips is encoded into the phase of the beat signal. The detection bandwidth of the photodiode is $50$ GHz and the electrical bandwidth of the fast oscilloscope is $36$ GHz. The beating signal is sampled at a rate of $80$ GSa/s. With these values, the beat signal is sufficiently well sampled for the proper quantitative determination of the band structure of the Floquet-Bloch lattice.

To obtain the dispersive band structure of the lattice, we perform the two-dimensional (2D) Fourier transform of the fringe pattern experimentally recorded and plotted in Fig. 2(a). The resulting 2D Fourier transform is computed numerically and shown in Fig.~2(b). The plotted spectrum spreads horizontally between $-15$ GHz and $+15$ GHz but the spectrum computed numerically spreads over a wider frequency span of $80$ GHz that is determined by the sampling rate of the oscilloscope. The vertical frequency range of the 2D spectrum is $6.5$ MHz, which corresponds to the mean free spectral range $\Delta \nu_{FSR}=1/\bar{T}$ of the double loop system. The 2D Fourier transform shown in Fig.~2(b) is composed of one vertical central band surrounded by two vertical side bands separated by $10.8$ GHz from the central zero-frequency component.

Figure 2(c) represents a zoomed view of the 2D Fourier spectrum plotted in Fig. 2(b) around $\sim 10.8$ GHz. It reveals that this region of the spectrum displays the double band structure characterizing the Floquet-Bloch lattice implemented in the double ring setup. The dashed square delimits the first quasi-mommentum and quasi-energy Brillouin zone. All eigenstates of both bands are excited by the initial input pulse. Fig. 2 shows that the band structure is straightforwardly determined only from the single-shot recording of the space-time pattern shown in Fig. 2(a) and the computation of its 2D Fourier transform. The measured shape of the bands agrees quantitatively with Eq.~\ref{eq:disp} for $\phi=0$ ($\cos\theta = (\cos Q - 1)/2$). Note that Fig.~2 shows the spatio-temporal evolution and the measured band structure for the U ring. Analogously, the same kind of measurements can be done for the output of the V ring, as discussed in Sec. \ref{Sec3}. 

\begin{figure}[htbp]
  \includegraphics[width=0.5\textwidth]{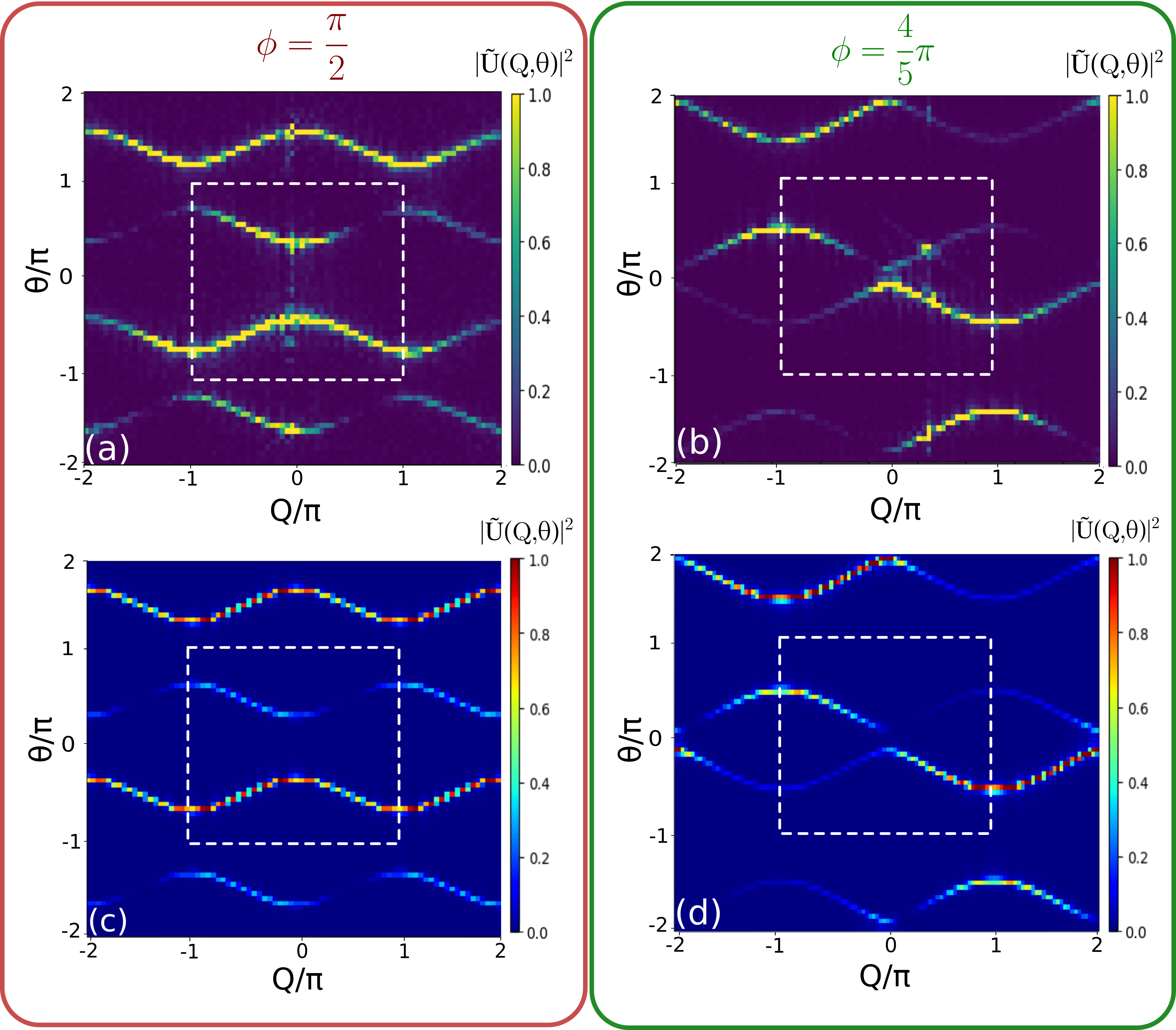}
  \caption{(a), (b) Experimental results showing the dispersive band structure of the photonic mesh lattice recorded with periodic square modulations of the phase with amplitudes $\phi_1= \pi/2$ (a) and $\phi_2=4\pi/5$ (b). (c), (d) Corresponding numerical simulations showing the band structure computed from the 2D Fourier transform of the impulse response of the system. Numerical simulations are made using Eqs. (\ref{U}), (\ref{V}) with $\Phi_1(m)=(-1)^m \, \pi/2$ (c) and $\Phi_2(m) = (-1)^m \, 4\pi/5$ (d). The squares in white dashed lines represent the first Brillouin zone. 
  }
\end{figure}

\begin{figure*}[!ht]
  \centering
  \includegraphics[width=1.\linewidth]{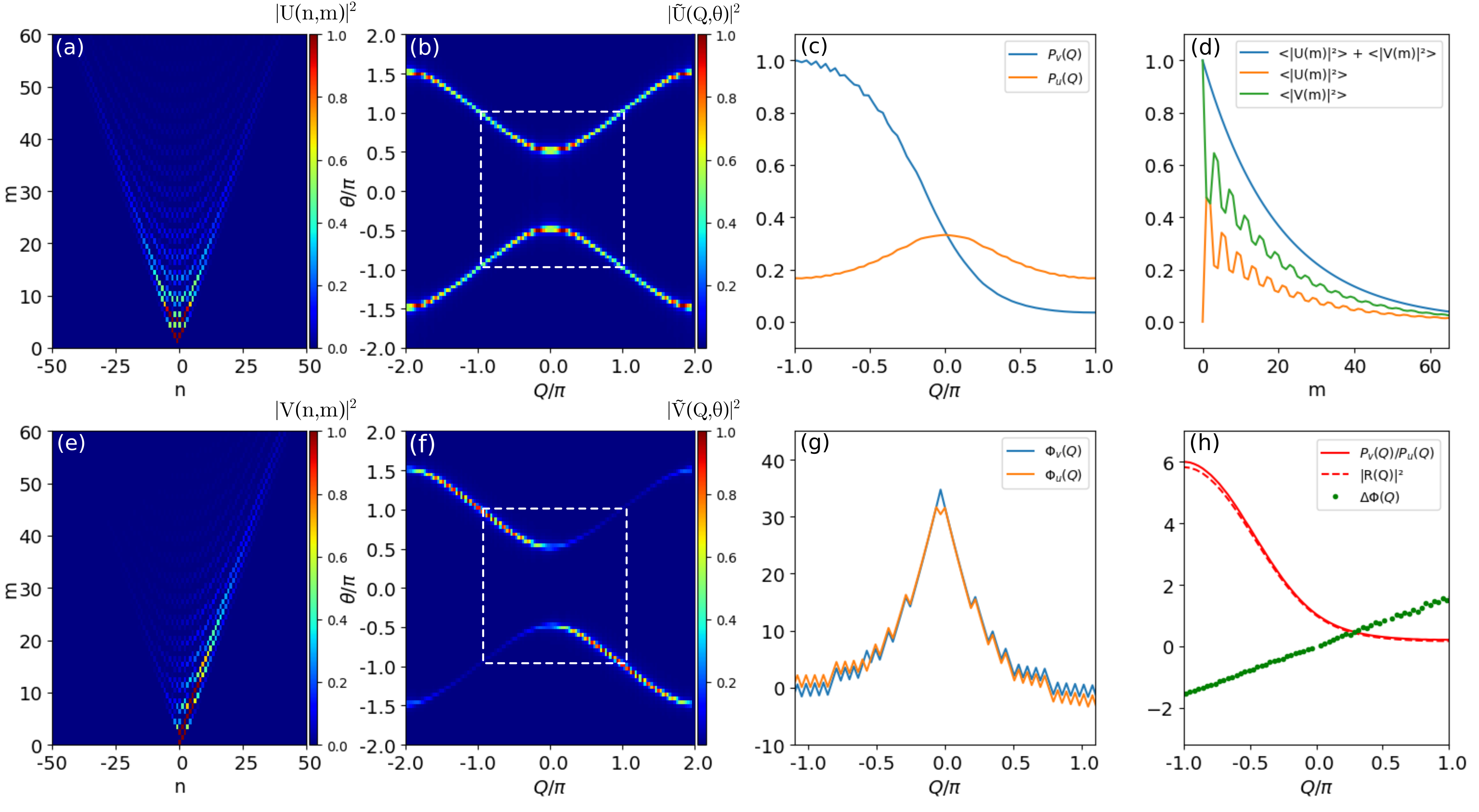}
  \caption{Numerical simulations of Eqs. (\ref{U}), (\ref{V}) showing the impulse responses of the two loops (a), (e) and their associated 2D Fourier transforms $|\tilde{U}(Q,\theta)|^2$ (b), $|\tilde{V}(Q,\theta)|^2$ (f). The squares in white dashed lines in (b) and (f) represent the first Brillouin zone. The simulations are made using a dissipation rate similar to the one measured in experiment ($\alpha = 0.07$) and no phase modulation ($\Phi(m)=0 \,\, \forall m$). (c) Spectral power distributions $P_{U}(Q)=|\tilde{U}(Q,\theta_{+}(Q))|^2$  (orange line) and $P_{V}(Q)=|\tilde{V}(Q,\theta_{+}(Q))|^2$ (blue line) measured along the upper spectral bands in (b) and (f). (d) Time evolution of the mean optical power in each loop : $ <|U(m)|^2> =  \mathop{\sum}_{n} |U(n,m)|^2$ (orange line), $ <|V(m)|^2> =  \mathop{\sum}_{n} |V(n,m)|^2$ (green line) and time evolution of the global power in the two loops $ <|U(m)|^2>+ <|V(m)|^2>$ (blue line). (g) Spectral phase distributions $\phi_V(Q)=Arg(\tilde{V}(Q,\theta_{+}(Q)))$ (blue line) and $\phi_U(Q)=Arg(\tilde{U}(Q,\theta_{+}(Q)))$ (orange line) measured along the upper spectral bands in (b) and (f). (h) Ratio between the spectral power distributions $P_V(Q)/P_U(Q)$ (red line) measured in the upper bands and evolution of the ratio $|R(Q)|^2$ (red dashed line) between the power of the Floquet-Bloch eigenmodes (see Eq. (\ref{R})). The spectral phase difference $\Delta \phi(Q)=\phi_V(Q)-\phi_U(Q)$ computed from the data shown in (g) is plotted in green line. It follows a simple linear relation given by $\Delta \phi(Q)=Q/2$ that complies with the fact that $Arg(R(Q))=Q/2$.
  }
\end{figure*}

In Fig. 2(c), the connection between the physical frequency $\nu_x$  (resp. $\nu_y$) measured on the horizontal (resp. vertical) axis and the Bloch momentum (resp. the quasi energy) is given by the following simple relation: $Q= 2 \pi (\nu_x-\nu_x^0)/  \Delta \nu_B$ (resp. $\theta=  4 \pi (\nu_y - \nu_y^0)/ \Delta \nu_{FSR}$). $\Delta \nu_B=1/\Delta T=443$ MHz represents the width of the Brillouin zone. $\nu_x^0$ represents the central frequency of the Brillouin zone in quasi-momentum dimension. In practice, its value slowly fluctuates from one recording to the other because the optical length of the fiber loop system is not actively stabilized with respect to the wavelength of the laser light. The slow and uncontrolled drift of $\nu_0^x$ arises from slow fluctuations of the difference in loop lengths $\Delta L$ on a timescale that typically falls in the second range. In practice the value of $\nu_0^x$ is ``manually'' selected from one recording to the other in such a way that the  band spectrum is symmetric with respect to the center of the Brillouin zone ($Q=0$). The same phenomenon occurs along the vertical frequency direction because the mean length $(2 L_1 +\Delta L)/2$ of the double loop system also fluctuates with respect to the laser wavelength. Consequently, from one recording to the next, there is a slow drift of the double band spectrum around the horizontal frequency axis. This drift effect is also corrected ``manually'' by adjusting $\nu_0^y$ in such a way that the band spectrum is symmetric with respect to the horizontal frequency axis ($\theta=0$). Active stabilisation of the ring length to an integer multiple of the wavelength of the laser used to inject the pulses would unambigously fix $\nu_0^x$ and $\nu_0^y$.

To explore other band structures, we now activate the PM inserted in the shorter loop (see Fig. 1(a)): the phase of the field in the shorter loop is modulated in time by a square signal oscillating between between $+\phi$ and $-\phi$ at a period equal to the mean round trip time $\bar{T}$ of light inside the loops. As shown in ref. \cite{Wimmer:17}, this modulation scheme permits to modify the band structure of the photonic lattice. Figures 3(a) and 3(b) show the band structure measured experimentally in the U ring for $\phi=\phi_1=\pi/2$ and for $\phi=\phi_2=4\pi/5$, respectively. Correspondingly, Figs. 3(c) and 3(d)) show the band structure numerically computed from the Fourier transform of the spatio-temporal evolution calculated using Eqs.~(\ref{U}) and~(\ref{V}) with a single pulse injected in the V ring as initial condition, like in the experiment. A good quantitative agreement is found between experiments and theory in terms of the shape and occupation of the bands for the selected values of $\phi$.

\section{Experimental determination of the eigenmode structure of the Floquet-Bloch lattice}\label{Sec3}

Measuring the impulse response of the photonic lattice by injecting a short pulse into the fiber system, we perform an excitation of the entire Brillouin zone that reveals the dispersive band structure of the lattice.  We will see now that it also permits to extract quantitative information on the structure of its eigenmodes. Assuming that the PM inserted in the short loop is inactive ($\Phi(m)=0$), the substitution of Eq. (\ref{eigenmodes}) into Eqs. (\ref{U}),(\ref{V}) provides not only the energy dispersion Eq.~(\ref{eq:disp}) 
but also the ratio $R$ between the complex amplitudes of the Floquet-Bloch eigenmodes in the two loops:
\begin{equation}\label{R}
  R=\frac{V}{U}=\frac{i}{\sqrt{2} \, \,  e^{iQ/2} \, \, \, e^{i\theta/2} -1}
\end{equation}

\begin{figure*}[!ht]
  \centering
  \includegraphics[width=1.\linewidth]{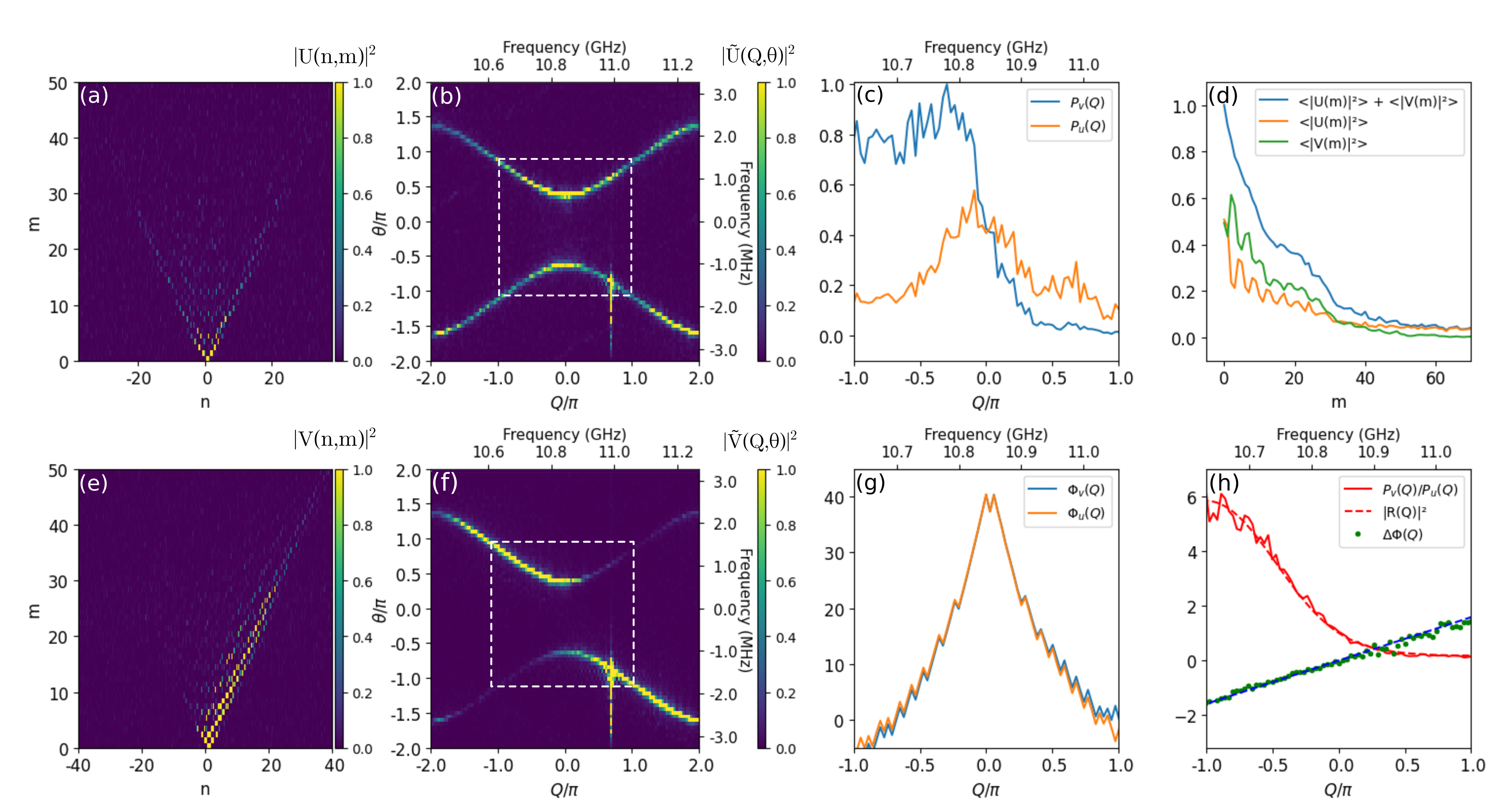}
  \caption{Same as in Fig. 4 but from experiments. 
  }
\end{figure*}

The meaning of Eq. (\ref{R}) can be better understood by analyzing numerical simulations of Eqs. (\ref{U}), (\ref{V}). Figure 4(a) and (e) show the computed spatio-temporal evolution of the lattice with $\Phi(m)=0$ and a single pulse injected in the V ring as initial condition: $v(n=0,m=0)=1$, $v(n \neq 0,m=0)=0$; $u(n,m=0)=0 \, \forall n$. In these simulations we have additionally added a term $-\frac{\alpha}{2} u_n^m$ and $-\frac{\alpha}{2} v_n^m$ in the right-hand side of Eqs. (\ref{U}) and (\ref{V}), respectively, to describe the fact that the losses in each loop are not perfectly well compensated by gain. As shown in Fig. 4(e) an asymmetric evolution of the light power distribution $|v(n,m)|^2$ is observed in the loop where the light pulse is initially injected (the V loop) while a symmetric one is observed at the output of the other loop (U loop), see Fig. 4(a). This asymmetric features between the injected and non-injected rings were already reported in the space-time domain in ref. \cite{Regensburger:11}. Fig. 4(b) (resp. Fig. 4(f)) represents the 2D Fourier power spectrum $|\tilde{U}(Q,\theta)|^2$ (resp. $|\tilde{V}(Q,\theta)|^2$) of the impulse response $u(n,m)$ (resp. $v(n,m)$) shown in Fig. 4(a) (resp. Fig. 4(e)). Asymmetric features observed in space-time domain are also found in the 2D Fourier spectra shown in Fig. 4(b) and in Fig. 4(f) for the U and V rings respectively.

From the 2D spectra shown in Fig. 4(b) and in Fig. 4(f), the spectral power distribution and the spectral phase distribution in the upper bands can be measured as a function of the Bloch momentum $Q$. The blue curve in Fig. 4(c) shows the spectral power distribution $P_V(Q)=|\tilde{V}(Q,\theta_{+})|^2$ measured for the V ring along the upper band of Fig.~4(f) whose dispersion is given by $\theta_{+}(Q)=+\arccos((\cos(Q)-1)/2)$, see Supplementary material for the description of the method used for the computation of the spectral power. The orange curve in Fig. 4(c) shows, equivalently, the spectral power distribution of the upper band for the U ring: $P_U(Q)=|\tilde{U}(Q,\theta_{+})|^2$. The ratio $P_V(Q,\theta_{+})/P_U(Q,\theta_{+})$ is displayed in full red line in Fig.~4(h). It nearly coincides with the dashed red curve, which displays $|R(Q)|^2$ computed from the analytical form given by Eq.~(\ref{R}).
The slight difference between solid and dashed red curves in Fig. 4(h) arises from the fact that numerical simulations have been made by incorporating dissipative effects ($\alpha \neq 0$) while the expression of $R$ is calculated for $\alpha=0$.

Analogously, by considering the argument of the 2D Fourier transform, we can compute the spectral phase distributions along the upper bands $\phi_V(Q)=Arg(\tilde{V}(Q,\theta_{+}(Q)))$ and $\phi_U(Q)=Arg(\tilde{U}(Q,\theta_{+}(Q)))$, for the V and U rings, respectively. In this case, for each value of $Q$, the phase is obtained at the value of the maximum spectral power density, see Supplementary material for details about the computation of the spectral phase. Figure 4(g) shows that the spectral phase in each band undergoes an excursion of $\sim 40$ radians over the entire Brillouin zone. Remarkably, the phase difference $\Delta \phi(Q)=\phi_V(Q)-\phi_U(Q)$ follows a simple linear evolution ($\Delta \phi(Q)=Q/2$) plotted in green in Fig. 4(h)), which fully complies with the evolution of the argument of $R(Q)$: $Arg(R(Q))=Q/2$, from Eq.~(\ref{R}). A similar treatment can be done for the lower bands. The analysis we have just described based on numerical simulations shows how the eigenmode structure of the photonic mesh lattice can be measured from the Fourier transform of the impulse response of the lattice. Let us note that the results synthesized in Fig. 4 show that the eigenstate structure determined from an analytical calculation where $\alpha$ is set to zero (Eq.~(\ref{R})) are robust to dissipative effects.

Figure 5 shows that all features revealed by numerical simulations reported in Fig. 4 are observed using our experimental methodology. Figures 5(b) and 5(f) display the symmetric and asymmetric 2D Fourier spectra that are computed from the impulse responses measured at the output of each fiber loop, see Fig. 5(a) and Fig. 5(e). Despite clear differences in the intensity distributions in the measured bands with respect to the bands computed from numerical simulations (compare Fig. 4(c) and Fig. 5(c)), the ratio $P_V(Q)/P_U(Q)$ between the measured spectral powers is very close to the theoretical curve over the entire Brillouin zone, see Fig. 5(h). The spectral phases $\phi_U(Q)$ and $\phi_V(Q)$ measured in the upper bands in Fig. 5(b) and Fig. 5(f) depict the same evolution as the one obtained from numerical simulations. Regarding the phase extracted of the upper bands extracted from the 2D Fourier transform protocol, Fig. 5(g) displays a large excursion in the phases $\phi_{V,U}(Q)$ over the Brillouin zone. Nevertheless, the phase difference $\Delta \phi(Q)=\phi_V(Q)-\phi_U(Q)$ follows a simple linear relation in good agreement with results obtained from the model (Eqs. (\ref{U})(\ref{V})): $\Delta \phi(Q)=Arg(R(Q))=Q/2$.

\section{Conclusion}\label{Sec4}

In this paper, we have reported experiments where the dispersive band structure characterizing a Floquet-Bloch lattice is measured in single shot over the entire Brillouin zone by Fourier transforming the impulse response of the lattice. In addition, our method provides the full and accurate characterization of the lattice eigenmode structure, i. e. the amplitudes and the phases of the Floquet-Bloch eigenvectors over the entire Brillouin zone. 

We believe that our experimental method will be useful not only for the accurate characterization of the linear dispersive properties of time-multiplexed photonic mesh lattices but also for the investigation of questions related to the influence of nonlinear effects on the propagation of Floquet-Bloch waves. A variety of nonlinear wave mixing phenomena at the origin of the broadening of spectral bands or of the nonlinear coupling between the Floquet-Bloch eigenmodes can be investigated using our experimental technique. In particular it could be useful to explore the phenomenon of modulation instability in nonlinear topological photonic system, as recently suggested in ref. \cite{Leykam:21}.

\section{Acknowledgements}
This work was supported by the H2020-FETFLAG project PhoQus (820392), European Research Council (865151) via the project EmergenTopo, the French government through the Programme Investissement d’Avenir (ISITE ULNE / ANR-16-IDEX-0004 ULNE) and IDEX-ISITE initiative 16-IDEX-0001 (CAP 20-25), managed by the Agence Nationale de la Recherche, the Labex CEMPI (ANR-11-LABX-0007), the CPER Photonics for Society P4S.


%

\pagebreak
\renewcommand{\theequation}{S\arabic{equation}}
\setcounter{equation}{0}
\setcounter{figure}{0}
\setcounter{section}{0}
\onecolumngrid

\newpage

\maketitle

\renewcommand{\theequation}{S\arabic{equation}}
\renewcommand{\thefigure}{S\arabic{figure}}

\begin{center}
\Large{\bf{Single shot measurement of the photonic band structure in a Floquet-Bloch lattice realised with coupled fiber rings: supplemental document}}
\end{center}

\begin{center}
\end{center}

\maketitle

\section{Reconstruction of the spatio-temporal diagrams from the time signal recorded in the experiments}

\begin{figure}[htbp]
\centering
\fbox{\includegraphics[width=0.6\linewidth]{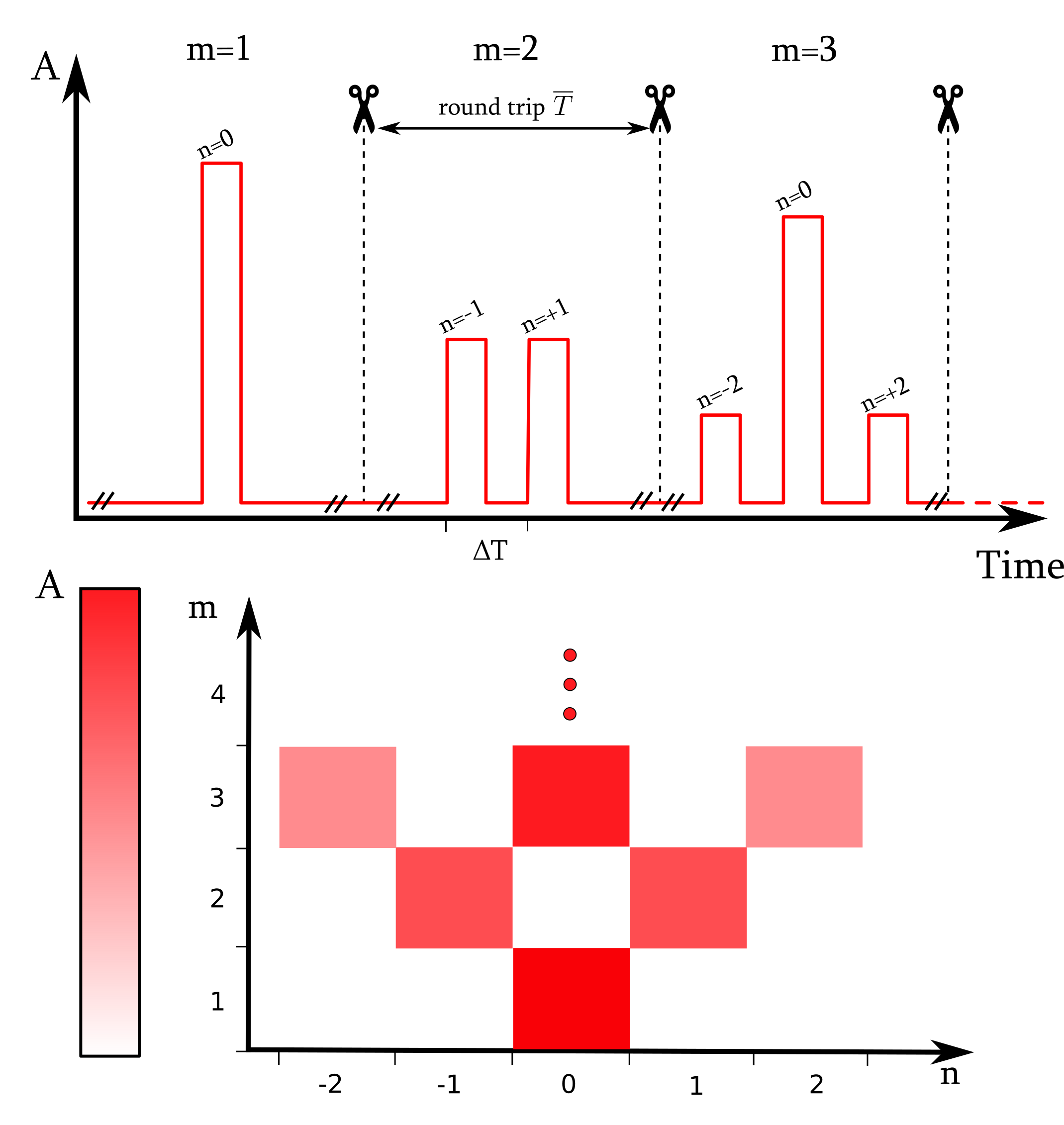}}
\caption{Schematic representation of the protocol used to reconstruct the space-time diagram showing the evolution of light pulses in the Floquet-Bloch lattice by starting from the time signal recorded by a photodiode at the output of the double loop system. The recorded signal is sliced in a sequence of multiple time windows having all a duration $\bar{T}$ equal to the mean round trip time of light in the double loop system. The space-time diagram is recomposed by concatenating the multiple time slices and by encoding the amplitude of the square pulses using a colormap.}
\label{fig1}
\end{figure}

In the experiment, an electrical signal changing in time is recorded at the output of each fiber loop by using fast photodiodes. This signal is composed of sets of square pulses having peak amplitudes that slowly decay in average at each round trip while also spreading out in time due to the slight length imbalance between the two fiber loops, see Fig. \ref{fig1}. Using other words the recorded signal evolves on a slow time scale $\bar{T}=(2L_1+\Delta L)/(2 v)$ determined by the mean round trip time of light in the fiber loop system and on a faster time scale $\Delta T=\Delta L/v$ determined by the difference length $\Delta L=L_2-L_1$ between the two loops. In the experiment, the two timescales $\bar{T} \simeq 152.5$ ns and $\Delta T= \simeq 2.26$ ns are well separated and it is appropriate to plot the dynamical evolution of light pulses using a representation where the slow evolution is decoupled from the fast one.

This is achieved by using a method that it is schematically shown in Fig. \ref{fig1}. The mean round trip time $\bar{T}$ is measured in an accurate way and the recorded time signal is sliced into a sequence of time windows having all a duration $\bar{T}$. The time segments obtained from this procedure are then concatenated in a space-time representation where the horizontal axis is associated with the fast time scale while the vertical axis is associated to the slow time scale. The spatiotemporal evolution of the light pulses in the photonic lattice is obtained in a last step where the amplitude of the square pulses is encoded by using a colormap.

\section{Measuring the spectral power distribution and the spectral phase distribution in the photonic band structure}

The method used in our paper to measure the photonic band structure of the Floquet-Bloch lattice consists in Fourier transforming the impulse response of the lattice. In addition to providing the shape of the dispersive bands, this method provides the complex amplitude of the Floquet-Bloch eigenmodes. Here we describe how the power and the phase of the eigenmodes are determined from the 2D band spectra that are computed by Fourier transforming the impulse response of the lattice. 

\begin{figure}[htbp]
\centering
\fbox{\includegraphics[width=1\linewidth]{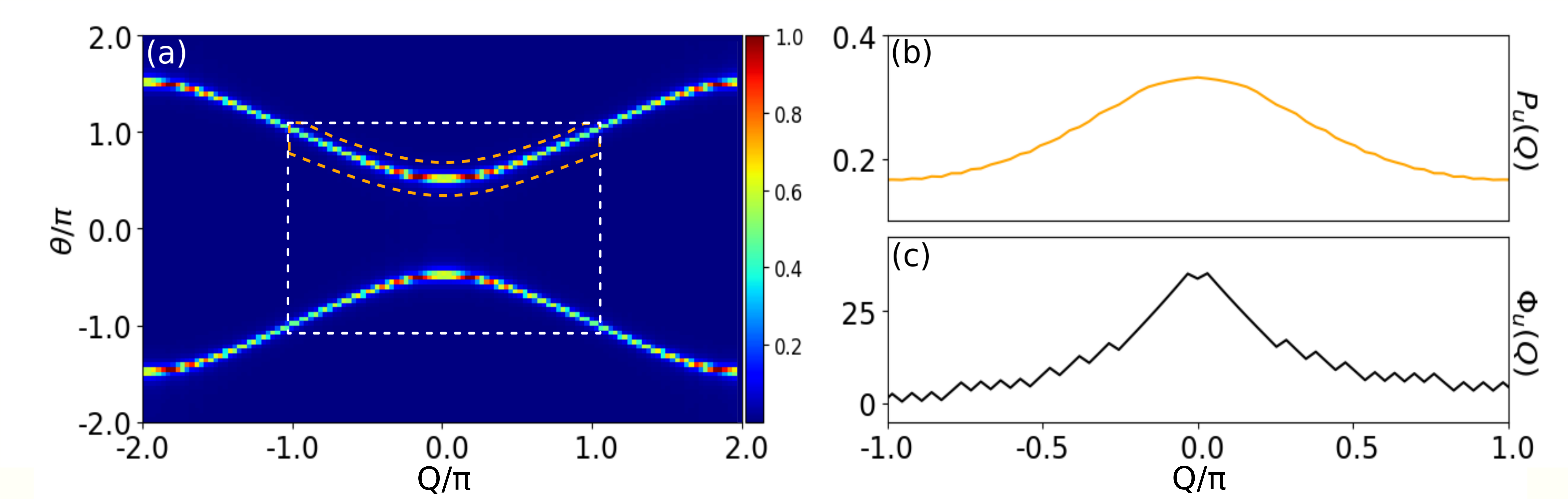}}
\caption{(a) Fourier power spectrum $\tilde{U}(Q,\theta)|^2$ of the impulse response of the Floquet-Bloch lattice for the symmetric U loop. The spectral power of the Floquet-Bloch eigenmodes shown in (b) is determined by integrating the spectral power density between the orange dashed lines using Eq. (\ref{P_U}) inside the first Brillouin zone delimited by white dashed lines. The spectral phase shown in (c) is the argument $\phi_U(Q)$ of the Fourier modes at the frequencies where the spectral power density is maximum.}
\label{fig2}
\end{figure}

The 2D Fourier power spectrum $|\tilde{U}(Q,\theta)|^2$ computed from the impulse response of the U loop is shown in Fig. \ref{fig2}(a). For any given value $Q_0$ of the Bloch momentum, the power $P_U(Q_0)$ of a Floquet-Bloch eigenmode is determined by integrating the spectral power density in a narrow spectral region centered around the upper dispersive band using the following expression 
\begin{equation}\label{P_U}
  P_U(Q_0)=\frac{1}{2 \Delta \theta} \int_{\theta_{+}(Q_0)-\Delta \theta}^{\theta_{+}(Q_0)+\Delta \theta} |\tilde{U}(Q_0,\theta)|^2 d\theta
\end{equation}
where $\theta_{+}(Q)=+\arccos((\cos(Q)-1)/2)$ provides the shape of the spectral band. Note that the knowledge of the analytical form of the function $\theta_+(Q)$ is not required and that the value $\theta_+(Q_0)$ simply represents the value of $\theta_+$ for which the spectral power density is maximum: $\theta_+(Q_0)=max(|\tilde{U}(Q_0,\theta)|^2)$. The value of $\Delta \theta$ is chosen in such a way that the average power given by Eq. (\ref{P_U}) represents a smooth function of $Q$, as shown in Fig. \ref{fig2}(b) ($\Delta \theta =0.1$ in Fig. \ref{fig2}).

To determine the spectral phase $\phi_U(Q)$ of the Floquet-Bloch eigenmodes, we first locate the points in the upper part of the 2D spectrum (Fig. \ref{fig2}(a)) where the spectral power density is maximum: $\theta_+(Q)=max(|\tilde{U}(Q,\theta)|^2$, $\theta>0)$. The spectral phase $\phi_U(Q)$ is simply computed as the argument of the Fourier modes at the points where the spectral power density is maximum: $\phi_U(Q)=Arg(\tilde{U}(Q,\theta_+(Q))$.

\end{document}